\documentclass[fleqn,10pt]{wlscirep}
\usepackage{bm}
\usepackage{xcolor}
\usepackage{float}
\floatstyle{boxed}
\usepackage{wrapfig}
\usepackage{array}
\usepackage{graphicx}
\usepackage{amsbsy}
\usepackage{amsbsy}
\usepackage{physics}

\newcommand\redline{\bgroup\markoverwith
    {\textcolor{red}{\rule[.5ex]{2pt}{1pt}}}\ULon}

\begin{document}
	
\title{Universal non-Markovianity Detection in Hybrid Open Quantum Systems}
	
\author[1,2,*]{Ji\v{r}\'{i} Svozil\'{i}k}
\author[1]{Ra\'{u}l Hidalgo-Sacoto}	
\author[2]{Ievgen I. Arkhipov}		

\affil[1]{School of Physical Sciences \& Nanotechnology, Yachay Tech University, 100119 Urcuqu\'i, Ecuador}
\affil[2]{Joint Laboratory of Optics of Palack\'{y}  University and Institute of
	Physics of CAS, Faculty of Science, Palack\'y  University, 17. listopadu
	12, 771 46 Olomouc, Czech Republic}
\affil[*]{jiri.svozilik@gmail.com}

\begin{abstract}
A universal characterization of non-Markovianity for any open hybrid quantum systems is presented. This formulation is based on the negativity volume of the generalized Wigner function, which serves as an indicator of the quantum correlations in any composite quantum systems. 
It is shown, that the proposed measure can be utilized for any single or multi-partite quantum system, containing any discrete or continuous variables.
To demonstrate its power in revealing non-Markovianity in such quantum systems, we additionally consider a few illustrative examples. 
\end{abstract}

\maketitle
	
\section{Introduction}
	
In Nature, any real quantum system can be significantly affected by its connection to surrounding ambient. This includes interchanges of information, excitations, or energy. 
Considering only the information interchange, if the information only flows from the system to the environment, the corresponding dynamics is said to be Markovian, as the environment is memory-less.
Whereas in any other case we deal with the non-Markovian environment \cite{breuer2002theory}. 
While for decades the Markovian dynamics has been in the central attention of investigators, a rapid development of open quantum systems strongly coupled to their environments showed major challenges.
In order to incorporate the back-action effects from environment to a system, several methods have been developed, e.g., the Nakajima-Zwanzig equation \cite{nakajima1958quantum,zwanzig1960ensemble} and  the time-convolutionless Master equation \cite{gorini1976n, gardiner2004quantum,ferialdi2016exact,nathan2020universal}. As examples of physical systems, where these models are appropriate, one can consider atoms in optical cavities \cite{cimmarusti2015environment}, superconducting circuits \cite{you2011atomic,xiang2013hybrid}, photonic-crystal cavity \cite{wu2010non}, etc.
	
In order to correctly characterize the dynamics of open quantum systems, many different measures of non-Markovianity have been introduced over time. This problem can be studied by taking different approaches and, obviously, by using different tools. The fundamental approach is based on the divisibility test of quantum maps $\Lambda(0,t)=\Lambda(t,s)\Lambda(s,0)$. That is, if the mapping can be divided at any time interval onto sub-intervals, this would imply a lack of history in the time evolution of a system~\cite{chruscinski2011measures}. The measure proposed by Breuer, Laine, amd Pillo \cite{breuer2009measure,urrego2018controlling} is based on the distinguishability of two quantum states quantified via their trace distance. When affected by any Markovian environment, their trace distance never increases with time. Another measure makes use of the entanglement evolution between the system and some ancillary state \cite{luo2012quantifying}, when connected to the environment. A somewhat different approach, based on the changes in volume space of accessible states for $N$-level systems and Gaussian states, was proposed in Ref.~\cite{lorenzo2013geometrical}. The application of correlation and a more general response functions in detecting non-Markovianity have been presented in Refs.~\cite{ali2015non,Strasberg2018,tamascelli2018nonperturbative,Santis2019,chaing2020nonmarkovian}. Just recently,  the proposal based on the quantized coherence was theoretically and experimentally examined \cite{wu2020detecting}. Moreover, the utility of neural networks and quantum computers in revealing non-Markovian dynamics have been shown in Ref.~\cite{Luchnikov2020} and Ref.~\cite{Headmarsden2020}, respectively.
For a more general overview on the non-Markovianity and its measures, we recommend to see Refs.~\cite{de2017dynamics,breuer2016review}. 

However, the previously mentioned measures are taking into account a particular structure of the system. To circumvent this, one would require a universal formulation for any kind of quantum states, which can be achieved, e.g., by using the quasi-probability distributions. The Wigner function (WF) \cite{wigner1932quantum} has become an indispensable tool in the both classical and quantum domains  \cite{alonso2011wigner}. It allows one to present any  quantum state using the real-valued function in position-and-momentum space.
This formulation is valid for systems with continuous variables (CVs). For discrete variables (DVs), the WF can be formulated	for discrete states projected to the basis of continuous functions \cite{wootters1987wigner}. Recently, in Refs.~\cite{tilma2016wigner,Rundle2019} the unifying approach based on the transformation kernels for each subsystem has been presented, which permits to define a generalized Wigner function (GWF) for an arbitrary complex quantum system, consisting of any DV or CV subsystems, so-called ``hybrid'' systems. The proposed framework not only unifies different variable domains, but also enables to visualize the quantum correlations in such systems~\cite{Davies2019,Rundle2020}.
	
In this contribution, we employ the definition of GWF for arbitrary quantum system to introduce a new measure of non-Markovianity. 
We demonstrate the power of the proposed measure to reveal non-Markovian dynamics in composite systems on two canonical examples, namely, on qubit-qubit and qubit-squeezed coherent states. 
The introduced measure is based on the changes of the negativity volume (NV) of GWF when the system is interacting with its environment\cite{arkhipov2018negativity,kenfack2004negativity}. As we show bellow, the Markovian
dynamics is reflected by a pure monotonic behaviour (decreasing) of the NV, as the system and the environment became more and more entangled. On the other side, the non-Markovian dynamics shows reduced (and also increasing) changes of NV. This feature offers us to quantify  a degree of non-Markovianity
based on the negativity volume of the GWF. Most importantly, such behaviour is observed for any CV, DV, or hybrid quantum systems, all being either single or multipartite. We also note that the concept of negativity volume of the standard Wigner function in a CV domain has been already utilized in detection of non-Markovianity in bosonic systems~\cite{Xiong2019}.

\section*{Hybrid state evolution}
A generic open quantum system dynamics can be described by a master equation: 
\begin{eqnarray}\label{ME}
		\frac{{\rm d}\hat{\rho}}{{\rm d}t}={\cal L}\hat\rho,
		\label{L1}
\end{eqnarray}
where $\cal L$ is, in general, a time-dependent Liouvillian superoperator,  and for a weakly coupled bosonic Markovian environment the Liouvillian $\cal L$ attains the Gorini-Kossakowski-Sudarshan-Lindblad form~\cite{Lindblad1976,GKS1976}. For a hybrid system, consisting of the discrete and continuous parts, the Liouvillian $\cal L$ can be decomposed  as follows
\begin{equation}
{\cal L}=\mathcal{L}_0+\mathcal{L}_d+\mathcal{L}_c,
\end{equation}
where
\begin{eqnarray}\label{Lg}
		\mathcal{L}_0&=&\frac{1}{i\hslash}\left[\hat{H},\cdot\right],\nonumber \\
		\mathcal{L}_{d}&=&\sum_{m}\gamma_{d,m}(t)\left(\hat{L}_{d,m}\cdot\hat{L}^{\dagger}_{d,m}-\frac{1}{2}\left\{ \hat{L}_{d,m}^{\dagger}\hat{L}_{d,m},\cdot\right\} \right), \nonumber \\
		\mathcal{L}_{c}&=&\sum_{n}\gamma_{c,n}(t)\left(\hat{L}_{c,n}\cdot\hat{L}^{\dagger}_{c,n}-\frac{1}{2}\left\{ \hat{L}_{c,n}^{\dagger}\hat{L}_{c,n},\cdot\right\} \right).
\end{eqnarray}
The Liouvillian superoperators ${\cal L}_0$,  ${\cal L}_d$, and  ${\cal L}_c$ describe the coherent, incoherent discrete and incoherent continuous system evolution, respectively. Such superoperators can induce either dissipation or amplification in reduced quantum systems~\cite{agarwal2012quantum}.  
$\hat L$ is a Lindblad operator with $\hat L^\dagger$ being its Hermitian conjugate;  the square $[\ ]$  and curl $\{\ \}$ brackets denote commutator and anticommutator, respectively;  the symbol $\cdot$ is a placeholder to indicate where an operator should be applied;
and the damping coefficients $\gamma(t)$ are, in general, considered to be time-dependent. {However, a weakly coupled bosonic non-Markovian environments can be as well introduced in Eq.(\ref{Lg}) as  the damping coefficients $\gamma(t)$ are allowed to flip their signs \cite{de2017dynamics}.}

The density matrix $\hat\rho$ of the hybrid state in Eq.(\ref{ME}) can be, thus, written as following~\cite{breuer2002theory,Rivas2011Book}  following~\cite{breuer2002theory,Rivas2011Book} 
\begin{equation}\label{r2}
\hat\rho=\sum\limits_{i=0}^{N^2-1}v_i\hat\rho_i,
\end{equation}
{ with $\vec{v}=(1,\vec{\boldsymbol{w}})$, where $w_i={\rm Tr}[\hat\rho'_i\hat\rho]$, and $\hat\rho'_i$($\hat\rho_i$) are left(right) Liouvillian eigenmatrices. We note that because the Liouvillian is a non-Hermitian superoperator, it can attain both left and right eigenmatrices. Eventually, for the time-dependend coefficients in Eq.(\ref{Lg}) is beneficial to express Eq.(\ref{r2}) in terms of more suitable states, whose evolution can be easily calculated.} The total Hilbert space of a system is  $N=N_d\times N_c$, where $N_d$ ($N_c$) is Hilbert space of discrete (continuous) part of a system.
Moreover, due to the CV part, the total Hilbert space is infinite, i.e., $N\to\infty$, since $N_c\to\infty$. Nevertheless,  it is always possible to make a truncation of Hilbert space at some finite $N$ without affecting much the very description of a system, though $N$ can still be large. Indeed, in real experiments one always deals with a finite number of photons in a system, thus, the truncation assumption, in practice, can be justified \cite{buvzek1992coherent,lund2008fault,miranowicz2014phase}.

 The generalized Wigner function $W$ for any hybrid system, is defined as follows~\cite{tilma2016wigner}:
\begin{equation}\label{W_def}
W={\rm Tr}\Big[\hat{\rho}\hat{\Delta}\Big],
\end{equation}
{where $\hat\Delta$ is a kernel operator, which is represented by a tensor product of kernels corresponding to DV and CV subsystems.
In particular, for a bipartite hybrid system composed of a qubit and photonic part, the kernel operator in Eq.~(\ref{W_def}) can be written as $\hat\Delta=\hat{\Delta}_q\otimes\hat{\Delta}_p$, 
where $\hat\Delta_{q}$ and $\hat\Delta_{p}$  are kernel operators corresponding to the qubit and photonic subsystems, respectively, and which read as\cite{tilma2016wigner}:}
\begin{equation}\label{5}  
\hat\Delta_q(\phi,\theta)=\frac{1}{2}\Big[\hat U\hat\Pi_q \hat U^{\dagger}\Big], \quad \hat\Delta_p(\beta)= \hat D\hat\Pi_p \hat D^{\dagger}.
\end{equation}
The operator $\hat U$ is a rotational operator in SU(2) algebra, namely $\hat U=e^{i\hat\sigma_3\phi}e^{i\hat\sigma_2\theta}e^{i\hat\sigma_3\Phi}$ with Pauli operators $\hat\sigma_i$, $i=1,2,3$, and angles $\phi,\Phi\in[0,2\pi]$, $\theta\in[0,\pi]$. $\hat\Pi_q= \hat{\mathbb I}_2-\sqrt{3}\hat\sigma_3$ is a parity operator of the qubit. The operator $\hat D$ is the displacement operator of the coherent state, i.e., $\hat D=e^{\hat a^{\dagger}\beta-\hat a\beta^*}$, where $\hat a$ ($\hat a^{\dagger}$) is annihilation (creation) boson operator. The corresponding bosonic parity operator reads as $\hat\Pi_p=e^{i\pi \hat a^{\dagger}\hat a}$.  The form of a kernel operator $\Delta=\hat{\Delta}_q\otimes\hat{\Delta}_p\otimes\cdots$ for a more complex hybrid states, e.g., consisting of $N$-level systems, can be found in Refs.~\cite{tilma2016wigner,Rundle2019}. The normalization condition $\int W{\rm d}\Omega=1$ is obtained by means of an appropriate integral measure ${\rm d}\Omega$, which is a product of 
normalized differential volume of SU(2) space of a qubit with a  Haar measure ${\rm d}\nu$~\cite{Tilma2002,Tilma2004,Rundle2019}. 

Now, by combining Eqs.~(\ref{r2}) and (\ref{W_def}), one arrives at the following expression for the Wigner function of the initial state $\hat\rho$:
\begin{equation}\label{W2}
W=\sum\limits_{i=0}^{N^2-1}v_iW_i,
\end{equation}
where $W_i$ are Wigner functions corresponding to $\hat\rho_i$ { obtained using Eq.(\ref{W_def})}.

The GWF negativity volume is found as \cite{kenfack2004negativity,arkhipov2018negativity}
\begin{equation}\label{NV}
\mathcal{N}=\frac{1}{2}\left[\int d\boldsymbol{\Omega}\left|W\left(\boldsymbol{\Omega}\right)\right|-1\right].
\end{equation}
The usefulness of using NV of the GWF lies in its ability to detect quantum correlations in hybrid systems~\cite{arkhipov2018negativity}, and thus its potential in identification of non-Markovian backflow of nonclassicality from environment to such systems. It is worth mentioning that the GWF can already exhibit negativity for single qubit states, which do not possess any quantum correlations~\cite{tilma2016wigner}, that is in a striking contrast to ordinary Wigner function, defined for continuous states, where the negative values signal the presence of nonclassicality~\cite{agarwal2012quantum}.

The divisibility condition for quantum maps \cite{chruscinski2011measures} mentioned earlier implies that the NV of the GWF remains a monotonic function of time and such the vector $\vec{v}$, in Eq.~(\ref{W2}), can only contract.
{Indeed, for the Markovian dynamics the components of the vector $\vec{v}$ exponentially decay.} This means that any deviation of system dynamics from, in general, the time-inhomogeneous Markovianity, will be reflected in the non-monotonical behaviour of the NV of the GWF. Therefore, any non-Markovian processes can induce the increase of the vector of states $\vec{v}$, namely $\vec{\boldsymbol{w}}$, and, thus, a sudden increase of the Wigner function negativity volume, which allows to detect a backflow of nonclassicality from an environment to a system. This is related to the flip of the sign in the decaying rates $\gamma<0$, will be detected by the change in the sign of the speed of the NV $\cal N$, which for Markovian dynamics is always negative, i.e, ${\rm d}{\cal N}_M/{\rm d}t<0$. Thus, in order to quantify the degree of non-Markovianity, we can define the following measure:
\begin{equation}
D_{\mathcal{N}}=1-\frac{\left|\int\frac{d\mathcal{N}}{dt}dt\right|}{\int\left|\frac{d\mathcal{N}}{dt}\right|dt},
\label{NM}
\end{equation}
which can vary in the range $D_{\mathcal{N}}\in[0,1]$, and, therefore, the Markovianity is determined whenever $D_{\mathcal{N}}=0$. 

We will demonstrate the ability of the measure $D_{\mathcal{N}}$, given in Eq.~(\ref{NM}), in its identification of the non-Markovian dynamics in hybrid open systems on several examples considered in the subsequent sections.

\section*{Environment models}  

	\begin{figure}[t]
	\centering
		\includegraphics[width=14cm]{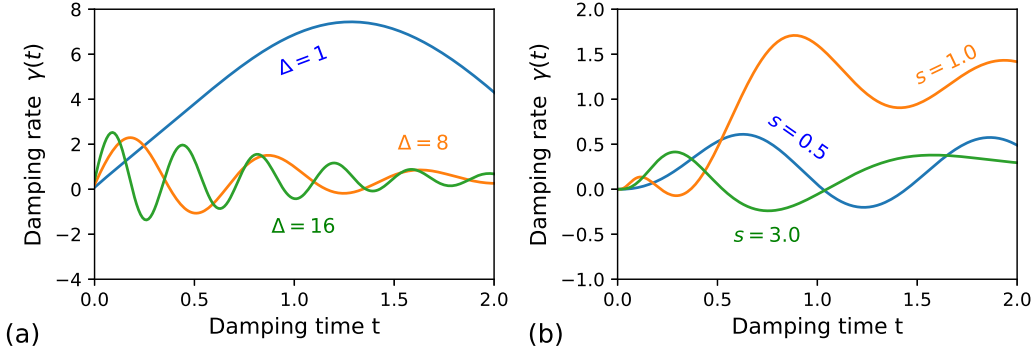}
		\caption{The damping functions for (a) the Lorentzian spectrum:
			$\delta_0=4.0$, $\lambda=1.0$, $\Delta=1.0$ (blue curve);
			$\delta_0=10.0$, $\lambda=2.0$, $\Delta=8.0$ (orange curve);
			$\delta_0=20.0$, $\lambda=2.0$, $\Delta=16.0$ (green curve); $c_2=c_4=1$;							
		and (b) the Ohmic-type spectrum:		
		  $s=0.5$, $\omega_0=5$, $\omega_c=10.0$, $\eta=0.01$ (blue curve);
		  $s=1.0$, $\omega_0=6$, $\omega_c=10.0$, $\eta=0.05$ (orange curve);
		  $s=3.0$, $\omega_0=5$, $\omega_c=2.54$, $\eta=0.05$ (green curve); $c_2=0$ $c_4=1$.}
		\label{Fig1}
	\end{figure}

In what follows, in our demonstration of the ability of the NV of the GWF to detect the non-Markovian dynamics of hybrid systems, without loss of generality, we will  focus on hybrid systems whose discrete part is formed by a qubit and the continuous part by a photonic field. 

As such, we consider the following incoherent Lindblad operators, in Eq.~(\ref{Lg}), for the qubit and photonic parts, respectively:
\begin{equation}\label{L}
		{\hat L}_q=\sqrt{\gamma_q(t)}\hat\sigma_z, \quad {\hat L}_{p,1}=\sqrt{\kappa_1(t)}\hat a, \quad {\hat L}_{p,2}=\sqrt{\kappa_2(t)}\hat n.
\end{equation}
\noindent Namely, the qubit decoherence is encoded by the Pauli matrix $\hat\sigma_z$. Whereas  the decoherence of the photonic part is represented by the amplitude damping $\hat{L}_{p,1}\propto \hat a$ and photon dephasing  $\hat L_{p,2}\propto \hat n$  processes, where $\hat a$ and $\hat n$ are the annihilation and photon number operators, respectively. The damping functions of the qubit and photonic parts, in Eq.~(\ref{L}), are denoted as $\gamma_q(t)$ and $\kappa_i(t)$, $i=1,2$, respectively.  
Also, we assume that the Hamiltonian $\hat H$ in Eq.~(\ref{Lg}) describes the free evolution of the qubit and photonic field. It attains the form $\hat H=\frac{\omega_{q0}}{2}\hat\sigma_z+\omega_{p0}\hat n$, where $\omega_{q0}$ and $\omega_{p0}$ is the qubit transition frequency and the photonic field frequency, respectively. We will work in the interaction picture, and, for simplicity, we set $\omega_{q0}=\omega_{p0}=\omega_0$, as we consider both decoherences acting separately in examined cases.

The non-Markovian evolution is introduced via the time convolution-less (TCL) projection operator techniques  employing up to the fourth-order correction terms representing the damping functions $\gamma_q(t)$, as well as $\kappa_1(t)$ and $\kappa_2(t)$, using the following expression originating in a perturbation expansion\cite{breuer2002theory}: 
\begin{eqnarray}\label{gam24}
    \gamma(t)&=&c_2\gamma_2(t)+c_4\gamma_4(t)\\
    \gamma_2(t)&=&Re\left[2\int_0^{t}dt_1f(t-t_1)\right]\\
    \gamma_4(t)&=&Re\left[2\int_0^{t}dt_1\int^{t_1}_{0}dt_2\int_{0}^{t_2}dt_3\left[f(t-t_2)f(t_1-t_3)+\right.\right.\nonumber\\
    &&\left.\left.+f(t-t_3)f(t_1-t_2)\right]\right],
\end{eqnarray}
	
\noindent where the time correlation function $f\left(\tau-\tau'\right)=\int_0^\infty d\omega J\left(\omega\right)e^{i\left(\omega_0-\omega\right)\left(\tau-\tau'\right)}$ {and $c_2$, $c_4$ are non-negative strength coefficients \cite{breuer2002theory}, which are employed to simply control the interaction, as they can be absorbed to coefficients for a reservoir.} Again, here, the form of the damping function $\gamma(t)$ formally embodies all the damping functions given in Eq.~(\ref{L}). That is, the functions $\gamma_q(t)$, $\kappa_1(t)$ and $\kappa_2(t)$ are all decomposed in the fashion presented in Eq.~(\ref{gam24}). As we are only interested in the evolution of NV, the frequency Lamb shifts induced by coupling to environments, which are time dependent, are ignored. One can also consider exactly solvable models of non-Markovian environments \cite{xiong2010exact,zhang2012general,xiong2015non}.{ For the sake of simplicity, we consider all damping functions in the form of Eq.(\ref{gam24}) and $\gamma_q(t)$=$\kappa_1(t)$=$\kappa_2(t)$ as we always select only one type of open quantum dynamics, given by  Eq.(\ref{Lg}), being enabled during a time evolution (for more details see \cite{agarwal2012quantum}). Nevertheless, all conclusions drawn in the following parts are applicable as well to any model of system-environment interactions, even with all open quantum dynamics combined together.}


Typical representative environments are described by the Lorentzian and Ohmic-type spectral functions. The first type of environment is related to an atom placed inside of a detuned optical cavity, which is represented by the Jaynes-Cummings model \cite{agarwal2012quantum}. Similar spectral function also describes a qubit, represented by a spin-1/2 system, coupled to a spin environment \cite{jing2018decoherence}. Such reservoir is characterized by the Lorentzian function \cite{el2019different}:

\begin{equation}
J_L\left(\omega\right)=\frac{1}{\pi}\frac{\delta_0\lambda^2}{\left(\omega_0-\Delta-\omega\right)^2+\lambda^2}.
\label{Eq:J1}
\end{equation}

{The parameter $\lambda$ is related to the relaxation rate and $\delta_0$ to the coupling.} The detuning of the cavity mode from the system transition frequency $\omega_0$ is denoted by $\Delta$. The corresponding correlation function for Eq.(\ref{Eq:J1}) is obtained as:

\begin{equation}
f_L(\tau-\tau')=\delta_0e^{-i\left(\tau-\tau'\right)\Delta-\lambda|\tau-\tau'|}
\end{equation}
Illustrative profiles of damping functions are shown in Fig.\ref{Fig1}a. The second type is represented by the Ohmic-type spectral function, which can imitate many different thermal baths and is applicable, for instance, to superconducting Josephson junctions\cite{makhlin2004dephasing} and nano-mechanical resonators \cite{seoanez2007dissipation}. The corresponding spectral function is given by \cite{gardiner2004quantum}:
	
\begin{equation}
J_O\left(\omega\right)=2\pi\eta\omega\left(\frac{\omega}{\omega_c}\right)^{s-1}e^{-\frac{\omega}{\omega_c}},
\label{Eq.J0}
\end{equation}

\noindent where $\omega_c$ is the critical wavelength of the environment spectra and  $\eta$ is the coupling strength. The parameter s denotes the Ohmic for $s=1$, sub-Ohmic for $0<s<1$, and super-Ohmic for $s>1$ cases. Three illustrative realizations are shown in Fig.\ref{Fig1}b. The corresponding correlation function is obtained as:

\begin{equation}
f_O(\tau-\tau')=\frac{2\eta\pi\Gamma\left(s+1\right) e^{i(\tau-\tau')\omega_0}}{\omega_c^{s-1}\left[i(\tau-\tau')+\frac{1}{\omega_c}\right]^{1+s}}.
\end{equation}
The symbol $\Gamma$ marks the gamma function.

		
\begin{figure}[t]
	\centering
	\includegraphics[width=14cm]{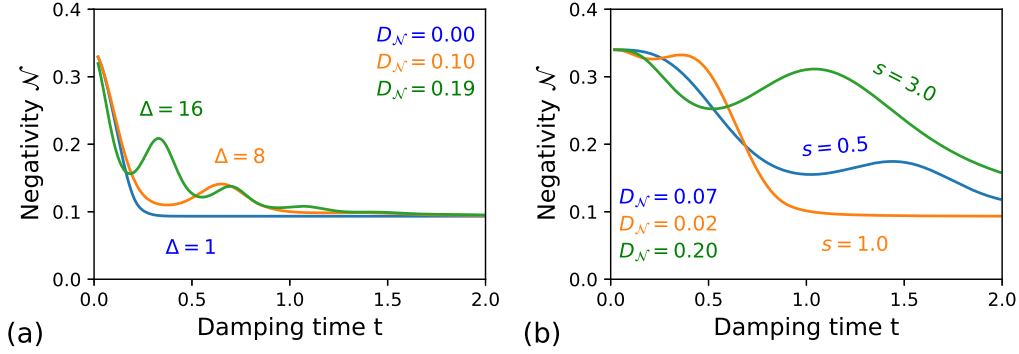}
	
	\caption{Negativity volume $\mathcal{N}$ for the qubit-qubit system coupled to (a) the Lorentzian environments: the blue curve $\Delta=1$ ($\delta_0=10$), the orange curve $\Delta=8$, the green curve $\Delta=16$; and (b) the Ohmic-type environments: the blue curve $s=0.5$, the orange curve $s=1$, the green curve $s=3$. The remaining parameters are the same as for the Fig.1.}
	\label{Fig:QQ}
\end{figure}		
		
\section*{Numerical simulations}

In this section, we numerically compute the NV of the GWF of the qubit-qubit and qubit--squeezed coherent states subjected to different non-Markovian environments. A photonic part in the second hybrid system is represented by a squeezed coherent state. For that we numerically integrate the master equation in (\ref{ME}) by applying the fourth-order Runge-Kutta method for different time-dependent damping functions, considered in the previous section.
The choice of the studied states is dictated by the frequent use of the qubit and squeezed coherent states in entanglement-based quantum information protocols utilizing the DV and CV states, respectively~\cite{karlsson2016non,chakraborty2019non}.
{By knowing the dynamics of a density matrix $\hat\rho(t)$, we then calculate the non-Markovianity measure $D_{\mathcal{N}}$, in Eq.~(\ref{NM}), for studied states to demonstrate its ability  to quantitatively capture the deviation of the time evolution of hybrid states from the ordinary Markovian dynamics, which is also reflected in the qualitative non-monotonic behaviour of the NV of the GWF.} 
		
\subsection{Qubit-qubit state}
Let us start, first, from the qubit-qubit state, written in the following form:
\begin{equation}\label{qqs}
|\Psi_1\rangle = \frac{1}{\sqrt{2}}\left(|00\rangle + |11\rangle\right).
\end{equation}
The state in Eq.~(\ref{qqs}) is one of the maximally entangled Bell states\cite{kwiat1995new}. Now, lets assume that the first qubit is coupled to a non-Markovian dephasing channel, modelled by the Lindblad operator $\hat L_q$ in Eq.~(\ref{L}).   
In Fig.\ref{Fig:QQ}a we plot the time evolution of the NV $\cal N$ of the GWF along with corresponding values of $D_{\mathcal{N}}$ for the qubit-qubit state for three different cases of Lorentzian environments, which have been depicted in Fig.~\ref{Fig1}a. 
The blue curve corresponds to the Markovian case, as the NV only gradually decreases in time. The other two curves (orange and green) exhibit  a partial revival in the negativity volumes indicating the non-Markovian evolution. The similar trends in the behaviour of the NV for the qubit-qubit state can be observed for the Ohmic-type environments shown in Fig.\ref{Fig:QQ}b. The different kinds of the Ohmic bath have been displayed in Fig.~\ref{Fig1}b. 

{Note also that in both Figs.~\ref{Fig:QQ}a and \ref{Fig:QQ}b, the NV tends to the value ${\cal N}=1/\sqrt{3}-1/2\approx0.08$ with $t\to\infty$, as the initial qubit-qubit entanglement eventually completely degrades. This is so-called {\it critical} value of the NV, namely, it is the maximal value of the NV which is generated by non-entangled, i.e., '{\it classical}' qubits. Above that critical value, the NV indicates the presence of the quantum correlations in the system~\cite{arkhipov2018negativity}.}

\begin{figure}[t]
	\centering
	\includegraphics[width=10cm]{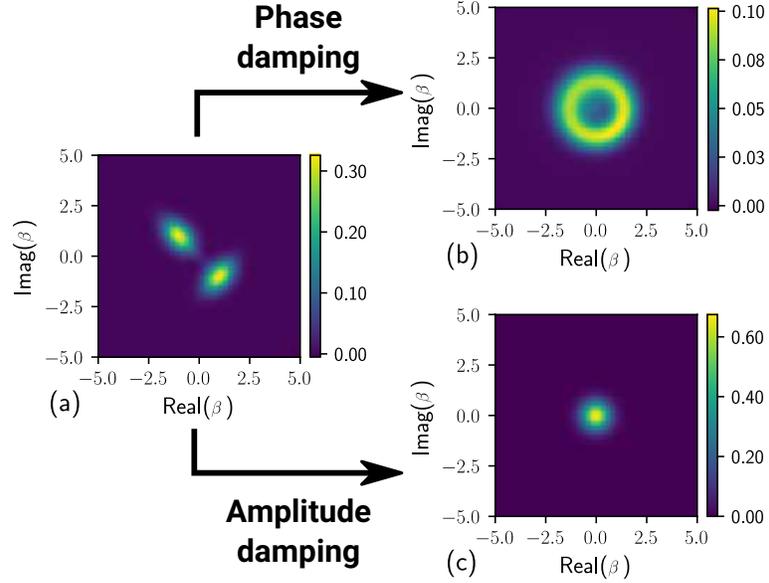}	
	
	\caption{The Wigner function for the photonic state given by Eq.(\ref{Eq:Phi2}) for the initial state  (a) and after it undergoes of the phase damping (b) and the amplitude damping (c). The state parameters are  $\alpha_1=-\alpha_2=1+i$, $r_1=r_2=0.15$, and $\phi_1=-\phi_2=1.5$ together with the Ohmic-type environment with $s=0.5$, $c_2=c_4=1$, and with (a) $\eta=0.005$ and (b) $\eta=0.01$. The remaining parameters are the same as for the Fig.1.}
	
	\label{Fig:QC_evol}
\end{figure}

\begin{figure}[t]
	\centering
	\includegraphics[width=14cm]{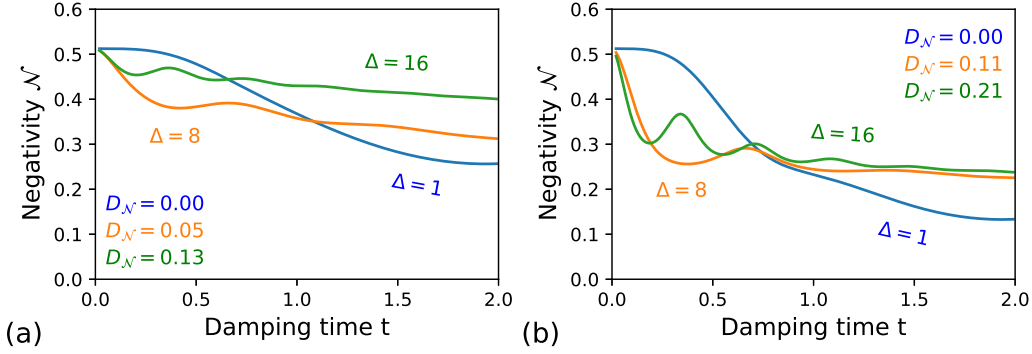}
	
	\caption{Negativity volume $\mathcal{N}$ for the qubit-photonic system, where the photonic part is coupled to the Lorentzian environment while undergoes of (a) phase and (b) amplitude damping. The blue curves correspond to $\Delta=1$, the orange curves to $\Delta=5$ and the green curves to $\Delta=16$. In (a) for $\Delta=1$ $\delta_0=2.5$, $\Delta=8$ $\delta_0=5$, $\Delta=16$ $\delta_0=5$; and in (b) for $\Delta=8$ $\delta_0=10$. The photonic state parameters are: $\alpha_1=-\alpha_2=1+i$, $r_1=r_2=0.15$, and $\phi_1=-\phi_2=1.5$. The remaining parameters of  are the same as for the Fig.~\ref{Fig1}.}
	\label{Fig:QC1}
\end{figure}
		
\subsection*{Qubit-Squeezed Coherent state}
{The second example, we would like to consider, is an entangled hybrid state\cite{andersen2015hybrid}, where a DV part is presented by a qubit and the CV part by a squeezed coherent state (SCS). This entangled hybrid state reads as}  
\begin{equation}
	|\Psi_{2}\rangle=\frac{1}{\sqrt{2}}\left(|0,\left(\xi_1, \alpha_1\right)\rangle+ |1, \left(\xi_2,\alpha_2\right)\rangle\right).
	\label{Eq:Phi2}
\end{equation}  
	
{The SCS is defined as
$|\xi_i,\alpha_i\rangle=\hat{D}\left(\alpha_i\right)\hat{S}\left(\xi_i\right)|0\rangle_c$,
where $\hat{D}(\alpha)=\exp\left(\alpha\hat a^{\dagger}-\alpha^*\hat a\right)$ is the displacement operator, the single-mode squeezing operator is $\hat{S}(\xi)=\exp\left[{\xi\frac{\hat{a}^{\dagger2}}{2}-\xi^*\frac{\hat{a}^{2}}{2}}\right]$, and $|0\rangle_c$ is vacuum state.}
{Moreover, we assume here that only the photonic part, i.e., SCS, is exposed to a non-Markovian environment.} {For our numerical calculation purposes, the Hilbert space for the photonic part can be effectively truncated as we consider small values of $|\alpha|$, which result in vanishing amplitudes for large photon numbers \cite{miranowicz2014phase}.}

\begin{figure}[t]
	\centering
	\includegraphics[width=14cm]{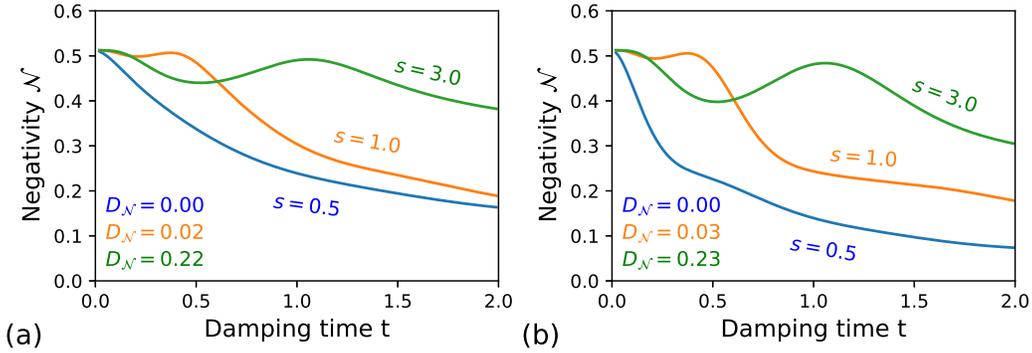}
	
	\caption{Negativity volume $\mathcal{N}$ for the qubit-photonic system, where the photonic part is coupled to the Ohmic-type environments while undergoes of (a) phase and (b) amplitude damping. The blue curves correspond to $s=0.5$, the orange curves to $s=1$ and the green curves to $s=3$. In (a) for s = 0.5 $\eta=0.005$. The remaining parameters are the same as for the Fig.~\ref{Fig:QC1}.}
	\label{Fig:QC2}
\end{figure}

To visualize the effect of the photon decoherence, we plot the GWF of the reduced photonic part for three different scenarios, namely, before and after the phase and amplitude damping in Figs.\ref{Fig:QC_evol}a, \ref{Fig:QC_evol}b and \ref{Fig:QC_evol}c, respectively.  The GWF $W_p(\beta)$ of the reduced photonic state is found by integrating out the degrees of freedom $\theta$ of the qubit state, that is 
$W_p(\beta)=\int d\theta W(\theta,\beta)$ (see Ref.~\cite{arkhipov2018negativity} for details).
In case of the phase damping [presented by the Lindblad operator $\hat L_{p,2}$ in Eq.~(\ref{L})], the reduced GWF of the initial SCS transforms as depicted in Fig.\ref{Fig:QC_evol}b. As one can see, the phase of the initial reduced SCS is smeared in time. Although the real amplitude remains unchanged, which is interpreted as a mixed state. Whereas in case of the amplitude damping [embodied by the operator $\hat L_{p,1}$ in Eq.~(\ref{L})], the GWF of the photonic state renders as shown in Fig.\ref{Fig:QC_evol}c, that is, it just degrades to the vacuum state $|0\rangle_c$.

The Lorentzian environment is presented in the Fig.\ref{Fig:QC1} for the both phase and amplitude damping channels and for varying parameters of environments. The blue curve in both plots marks the Markovian case, and the yellow and green curves are related to the non-Markovian case. One can see the related degrees of non-Markovianity, given by Eq.(\ref{NM}), in corners of each plot. Additionally, the main distinctiveness between curves in Fig.\ref{Fig:QC1} is the minimum $\mathcal{N}$ achieved in each case. In the case of amplitude damping Fig.\ref{Fig:QC1}b, the Wigner function reaches the minimal $\mathcal{N}$ in the Markovian dynamics, in a given time interval, whereas other curves are approaching this value only slowly. This means that the squeezed coherent state turns into the vacuum state $|0\rangle_c$ represented in Fig. \ref{Fig:QC_evol}c and a source of negativity is hidden in the qubit part. For the phase dumping, the resulting state, which is a mixed state, contains still more correlations than the previous case, which is reflected by higher values of $\mathcal{N}$, see Fig.\ref{Fig:QC_evol}b. A quite similar behavior can be observed  as well in Fig.\ref{Fig:QC2} for the Ohmic-type environments.

\section*{Conclusion}
	
We have introduced a new measure of non-Markovianity based on the negativity volume of the generalized Wigner function, which can be applied to any quantum composite system, i.e., hybrid system.
We have illustrated its power to reveal non-Markovian dynamics by considering a few canonical examples. Namely, we have applied it to the qubit-qubit and qubit--squeezed-coherent hybrid states. Such composite systems have been exposed to different bosonic environments. 
{ Despite the fact that the photonic part was only considered to be a single-mode field, all conclusions remain the same for the multi-mode photonic fields as only computation demands increase.}

Most importantly, our proposed measure is state-{\it independent} and, therefore, is {\it universal}, as it is solely based on the GWF defined for any hybrid state. Moreover, the experimental advances~\cite{Eichler2012} allow us to expect that the experimental measurements of the GWF will become a standard tool in characterizing complex composite quantum systems and their dynamics in the near future. Additionally, the knowledge of the GWF offers the opportunity to perform not only quantitative but also qualitative analysis of the time evolution of any quantum hybrid systems~\cite{Davies2019,Rundle2020}.

We note, however, that the knowledge of the NV of the GWF, in general, requires an implementation of a state tomography, which can be source demanding. Nonetheless, the task can be substantially simplified if the initial state is known~\cite{masse2020}. In this case, one can focus on detecting a specific region of the GWF, which demonstrates the largest concentration of the NV in a system, and, especially, when such a region exhibits an interference-like pattern, which indicates the region where the quantum correlations are revealed the most. This allows not only to probe the quantum correlations in a hybrid system with limited sources, but, consequently, also the non-Markovianity of its environment.

We, thus, believe that our results will ignite further developments in devising new experimental schemes allowing to reconstruct a GWF of any hybrid state, whose negativity volume can serve as a universal tool in detecting the nonclassicality and non-Markovianity in composite hybrid systems.

	\bibliographystyle{naturemag}
	\bibliography{main_final}
	
\noindent {\bf Acknowledgment}

J.S. and R.H.S. acknowledge the support from the School of Physical Sciences \& Nanotechnology of the Yachay Tech University. J.S. and I. A. thank the project CZ.02.1.01\/0.0\/0.0\/16\_019\/0000754 of the Ministry of
Education, Youth and Sports of the Czech Republic. I.A. also thanks the Grant Agency of the Czech
Republic (Project No.~18-08874S).

\noindent {\bf Contributions}

J.S. conceived the idea. J.S. and I.A. developed the theory. J.S. and R.H.S performed the analytical and  numerical calculations. All authors contributed to writing the manuscript.

\noindent{\bf Competing interests} 

The authors declare no competing  interests.
	
\end{document}